\documentclass[journal,transmag]{IEEEtran}
\ifCLASSINFOpdf
\else
\fi
\usepackage{amsmath}
\usepackage{amssymb}
\usepackage{graphicx}
\usepackage{epsfig}
\usepackage{url}
\usepackage{hyperref}
\newtheorem{Theorem}{Theorem}

\usepackage{color,rotating}
  \newcommand{\bT}{\mbox{\boldmath $T$}}

\newcommand{\cqfd}{\mbox{}\nolinebreak\hfill\rule{2mm}{2mm}\medbreak\par}
\begin{document}
\title{Multilevel Pricing Schemes in a Deregulated Wireless Network Market}


\author{\IEEEauthorblockN{Andrey Garnaev\IEEEauthorrefmark{1},
Yezekael Hayel\IEEEauthorrefmark{2}, and Eitan Altman
\IEEEauthorrefmark{3}}\\
\IEEEauthorblockA{\IEEEauthorrefmark{1} Saint Petersburg State
University,
St Petersburg, Russia, email: garnaev@yahoo.com}\\
\IEEEauthorblockA{  \IEEEauthorrefmark{2}LIA, University of Avignon,
Avignon, France, email: yezekael.hayel@univ-avignon.fr}\\
\IEEEauthorblockA{\IEEEauthorrefmark{3}INRIA, Sophia-Antipolis, France, email: eitan.altman@inria.fr}}

% REMEMBER HOWEVER: After having produced the .bbl file,

% and prior to final submission,
% you need to 'insert'  your .bbl file into your source .tex file so as to provide
% ONE 'self-contained' source file.
%
% Questions regarding SIGS should be sent to
% Adrienne Griscti ---> griscti@acm.org
%
% Questions/suggestions regarding the guidelines, .tex and .cls files, etc. to
% Gerald Murray ---> murray@hq.acm.org
%
% For tracking purposes - this is V3.1SP - APRIL 2009

\maketitle

\begin{abstract}
Typically the cost of a product, a good or a service has many
components. Those components come from different complex steps in
the supply chain of the product from sourcing to distribution.
This economic point of view also takes place in the determination
of goods and services in wireless networks.  Indeed, before
transmitting customer data, a network operator has to lease some
frequency range from a spectrum owner and also has to establish
agreements with electricity suppliers. The goal of this paper is
to compare two pricing schemes, namely a power-based and a flat
rate, and give a possible explanation why flat rate pricing
schemes are more common than power based pricing ones in a
deregulated wireless market. We suggest a hierarchical
game-theoretical model  of a three level supply chain: the end
users, the service provider and the spectrum owner. The end users
intend to transmit data on a wireless network. The amount of
traffic sent by the end users depends on the available frequency
bandwidth as well as the price they have to pay for their
transmission. A natural question arises for the service provider:
how to design an efficient pricing scheme in order to maximize his
profit. Moreover he has to take into account the lease charge he
has to pay to the spectrum owner and how many frequency bandwidth
to rent. The spectrum owner itself also looks for maximizing its
profit and has to determine the lease price to the service
provider. The equilibrium at each level of our supply chain model
are established and several properties are investigated. In
particular, in the case of a power-based pricing scheme, the
service provider and the spectrum owner tend to share the gross
provider profit. Whereas, considering the flat rate pricing
scheme, if the end users are going to exploit the network
intensively, then the tariffs of the suppliers (spectrum owner and
service provider) explode. \footnote[1]{This is  the last draft version of the paper. Revised version of the paper accepted by ValueTools 2013 can be found in Proceedings of the 7th International Conference on Performance Evaluation Methodologies and Tools (ValueTools '13), December 10-12, 2013, Turin, Italy}
\end{abstract}

%A category including the fourth, optional field follows...
%\category{D.2.8}{Software Engineering}{Metrics}[complexity measures, performance measures]

\begin{IEEEkeywords}
End user, Service provider, Spectrum owner, Flat rate pricing scheme, Power based pricing scheme, Pricing mechanism, Spectrum supply chain, Stackelberg equilibrium
\end{IEEEkeywords}

\section{Introduction}
\noindent Optimal pricing mechanisms have been widely studied in
networking to control the usage of a sparse resource like
frequency bandwidth. In the present work we investigate the
following question: what is the impact of the provider pricing
mechanism on the decision of each agent in a spectrum supply
chain. Often the organization of wireless networks is based on the
interaction among several economic entities. For instance, end
users pay for a given data rate. Service providers have some costs
associated with the wireless network services such as electricity
cost, office renting cost, frequency license, etc.

As stated in \cite{UN01}, regulation has been traditionally used
in several markets like infrastructure facilities and services.
 Often after deregulation there is
increased competition that in many cases benefits consumers. In
\cite{UN01}, the author describes the scope for competition in the
major utilities like Electricity, Gas, Railways,
Telecommunications, etc. The author shows that the competition is
feasible and desirable specifically in telecommunication market.
Moreover, in US, collusive behaviours, allowing cartel firms,
through a distortion of free market forces, to achieve and share
monopolistic profits, are forbidden and persecuted by (not by
chance called) ''anti-trust'' legislations \cite{Sama08}. That is
why in our paper we focus on a deregulation model between the
infrastructure provider and the service provider.

To understand the spectrum supply chain problem, we propose to
model the system as a multilevel economic game among various
agents. Specifically, we consider the following players; the end
users, the service provider and the spectrum owner. The spectrum
owner has the resource, the spectrum here, and is considered as
the top leader in the hierarchy. The service provider rents a
quantity of the resource in order to propose services to the end
users. The multilevel game approach means that the decision of the
spectrum owner is followed by an optimal decision of the service
provider, which in turns is followed by an optimal decision of the
end users.

Depending on the kind of wireless technology used by the service
provider, models of interaction between end users are different.
In a first approach, we consider an interference-free model where
each end user has his own frequency band to transmit his data.
Such kind of phenomenon appears considering Orthogonal Frequency
Division Multiplexing (OFDM) technologies \cite{Garg07}. In an
alternative approach, we consider a noncooperative power control
game between the end users which are interacting together through
their signal to interference and noise ratio (SNR). This second
model of interaction between the end users corresponds to another
wireless technology like code division multiple access (CDMA)
\cite{Garg07}.

We focus in this paper on a wireless market in which the spectrum
owner, called also the infrastructure provider, sells frequency
bandwidth to the service provider. This situation is typical in
economic models of Cognitive Radio Networks \cite{CRN09}. At a
lower level of the spectrum supply chain, the service provider
sells services to the end users. The user has his/her utility
function and pays according to the transmission rate tariff. The
service provider can be seen as a Mobile Virtual Network Operator
(MVNO) who obtains spectrum through long-term contracts with
spectrum owners or a Mobile Network Operator (MNO). The MVNO
resells the spectrum to the end users. MVNOs are mobile service
offerings from organizations which typically neither own licensed
mobile spectrum nor operate a physical base station network and
backhaul. Of course, there are exceptions to the rule but the
essential factor is that a MVNO cooperates with a MNO for network
access. MNOs are traditional mobile companies such as Orange,
SingTel and Vodafone \cite{MVNO10}. As stated in
\cite{Dewenter06,Duan11}, it is more efficient for a spectrum
owner to hire a MVNO to serve the end users because the MVNO can
have a better understanding of local population and user's
demand.Therefore, a natural question arises for the provider:
which tariff the service provider has to assign to obtain the
maximal pure profit, i.e. the difference between how much he
obtains from the end users and how much he has to pay for the
licensed frequency bandwidth to the spectrum owner. The spectrum
owner who rents part of his spectrum to a MVNO, in turn looks for
the frequency bandwidth tariff which can bring to him maximal
profit. We focus in our model on a deregulated spectrum market
where the resource and the services are not controlled by the same
provider, nor by providers which are colluding.

We would like to note that a Stackelberg game approach is very
popular among researchers dealing with pricing in networks (see,
for example, \cite{AAAA2008,  BS2002b, Bernstein04, Duan10, AAG2010b, AAG2010c,    MailleGT11,MSW2009, NH07,  GG2008}).
In \cite{Duan11}, the authors propose a four stage Stackelberg
model for the study of spectrum market pricing mechanism with
Cognitive MVNO who resells spectrum to secondary users. The
authors show some threshold structure of the network equilibrium
and fair spectrum allocations to secondary users. The first main
difference between the scenario suggested in the present paper and
those mentioned above is that we deal with three levels hierarchy
users-provider-spectrum owner. Second, we consider different
pricing mechanism for the service provider. Finally, we
investigate the impact of the interference model between end users
on the equilibrium. The major contributions of this paper are
described in the following items.
\begin{description}
\item[(i)] The power-based pricing mechanism implies the same profit for the service provider and the spectrum owner.
\item[(ii)] The flat rate pricing mechanism induces a higher profit for the service provider compared to the profit of the spectrum
owner. It gives  a possible explanation  to why flat rate pricing
schemes are more common than power based pricing ones.

\item[(iii)] A high SINR regime is always desirable in wireless systems to provide quality of service to the end users. Then, assuming this property, we get a simpler closed-form expressions of the equilibrium. Finally, we obtain that the power-based pricing mechanism,
assuming high SINR, leads to zero profit for the service provider.
Whereas the flat rate pricing mechanism induces non-zero profit
for the service provider.
\end{description}
The paper is organized as follows. We present first in section
\ref{sec:model}, the multilevel economic model with the decision
set of each agent at each level. In section \ref{general} we give
explicit formulations of the equilibrium in the general wireless
context. By assuming a high SINR regime, in section \ref{special},
we obtain a closed-form equilibrium and conclude saying that the
service provider has to consider the flat rate pricing mechanism
instead of the power-based one, in order to maximize his profit.
Finally, in Section~\ref{sec:conc} and Appendix discussions and sketch of proofs for
several theorems  are offered.

\section{Multilevel hierarchical pricing scheme}\label{sec:model}

We introduce a multilevel hierarchical game with the following
decision makers: end users, service providers and spectrum owners.
The system considered in this paper is a particular multi-level
hierarchical supply chain in a wireless network. In fact we
consider first  that at the lower level, with the longest time
scale, each end user determines his transmit power in order to
maximize selfishly his utility function. Second, a service
provider, at the middle level and intermediate time scale, rents
some frequencies from the spectrum owner in order to provide
services to the users. Then the service provider determines
optimally the quantity of bandwidth to rent to end users and the
tariff end user should pay for their services. Third, the spectrum
owner, at a longer level and shorter time scale, manages the
spectrum and determines the price of a unit of bandwidth rent by
the service provider. In terms of timescale, our system considers
three different timescales. The shorter one is the spectrum owner
one who decides the tariff per unit of frequency bandwidth the
service provider has to pay for. In telecommunication market, this
timescale can be like several years. A second longer timescale is
the service provider's one, which determines the quantity of
bandwidth to rent and the tariff each end users have to pay for
their services. This second timescale is the order of a year.
Finally, the longest timescale is the end users one who choose
their transmit power with a timescale of some months or days or
minutes, depending on the wireless technology used. Note that this
power control management should be implemented into communication
protocols in end-users devices. The main objective of our model is
to study this multilevel supply chain in wireless
telecommunication market, not to necessary obtain any practical
implementation or software, but to understand the behavior of the
different decision makers in such a complex economic system.

The three-level optimization problem can be described more
formally as follows from the spectrum owner (top-level) to the end
users (low-level).

(i) {\it The spectrum owner maximizes his revenue} $v_A(W,C_W)=C_W W$ depending on the tariff $C_W$ per unit of frequency bandwidth assigned to the service provider, i.e.
    \begin{eqnarray}
    \max_{C_W}v_A(W,C_W)=\max_{C_W}C_WW,\label{auth_payoff}
    \end{eqnarray}
    where $W$ is the quantity of bandwidth chosen by the servi ce provider.

(ii) {\it In order to maximize his payoff} $v_P(C_P,W,C_W)$, the service provider determines the quantity of bandwidth $W$ to license from the spectrum owner and the tariff $C_P$ the end users have to pay for their services i.e.
        \begin{eqnarray}
        \max_{W,C_P} v_P(C_P,W,C_W)= \max_{W,C_P}\left\{C_P\sum_{i=1}^{n}\mu(T_i)-C_W W\right\},\label{provider_payoff}
         \end{eqnarray}
        where the function $\mu(\cdot)$ is the pricing scheme used by the service provider, which depends on user's $i$ transmission power $T_i$ and $n$ is the number of end users.

(iii) {\it Finally,} each end user $i$ determines his transmission power $T_i$ in order to maximize his net utility $u_i(\bT,C_P)$ with $\bT=(T_1,\ldots,T_n)$ which is the difference between
        the throughput and the price imposed by the service provider, i.e.
              \begin{eqnarray}
            \max_{T_i} u_i(\bT,C_P)= \max_{T_i}\left\{W\ln\left(1+\gamma_i(\bT)\right)-C_P\mu(T_i)\right\},\label{endusers_payoff}
        \end{eqnarray}
where $\gamma_i(\bT)$ is the signal to noise ratio
(SNR) on user's $i$ signal. This function determines the quality
of the signal received and then can used to approximate the
capacity of a wireless link with bandwidth $W$ based on the
Shannon formula \cite{shannon48,RSV2009}. This utility function is
well-known in the literature on pricing schemes in wireless
systems like in \cite{AAAA2008}.

In order to compute the solutions of this multilevel game, we use
a step-by-step approach from the low-level optimization problem
(the end users) to the high-level one (the spectrum owner). The
solution of this three-level game is a hierarchical (Stackelberg) equilibrium  \cite{fudenberg91}.

{\it In the first step}, for a fixed usage tariff $C_P$
and bandwidth $W$, determined by the service provider, the $n$
users compete by trying to maximize selfishly their own payoff. We
deal here with a homogeneous network, namely, we assume that all
the users are in the same wireless conditions (so, they have the
same restriction on power and the same fading gains). Each of them
can be considered as an average user. Homogeneous network suits
considering the network where average users operate and since we
try to give an explanation why flat rating is more common we can
restrict ourself to such network. In our future work we are going
to investigate what impact on pricing schemes different types of
user's groups can produce, say, employing different applications or have different objectives
\cite{GTK2012,GTK2013}.

It is natural when studying a power control problem, to consider
that a strategy of a user $i$ is the transmitted power $T_i\in
[0,\overline{T}]$ with $\overline{T}$ is the maximal (average)
power a user can apply.
%The payoff of user $i$ is
%given as follows:
%\begin{eqnarray}
%u_{i}(T_1,\ldots,T_n)=W\ln\left(1+\frac{\displaystyle L h
%T_i}{\displaystyle W\sigma^2+h\sum_{j=1,j\not=i}^n T_j}\right)-C_P\mu(T_i), \label{user_payoff}
%\end{eqnarray}
%\normalsize \noindent where $\sigma^2$ is the background noise, $h$ is the fading channel gain and $L$ is the crosstalk coefficient.

 {\it In the second step} of the multilevel game, the
service provider, knowing that the end users will act as it was
described in the fist step, chooses for the optimal tariff $C_P$
and how much frequency bandwidth $W$ to license from the spectrum
owner. The service provider payoff $v_P$ is the difference between
how much he earns from selling services to the users and how much
he has to pay for the licensed frequency bandwidth.
%Then, the
%provider's payoff is given as follows:
%\begin{eqnarray}
%v_P(C_P,W)=C_P\sum_{i=1}^n \mu(T_i)-C_W W, \label{prov_payoff}
%\end{eqnarray}
%where $C_W$ is the tariff per unit of bandwidth assigned by the authority
%at the third step.

 {\it In the third step} of the multilevel game, the
spectrum owner selects the optimal tariff $C_W$ it has to assign
to get the maximal profit.
%Thus, the authority gain $v_A$ is
%given by
%\begin{eqnarray}
%v_A(C_W)=C_W W. \label{auth_payoff}
%\end{eqnarray}

The aim of the paper is to study this multilevel economic model
with two pricing schemes for the service provider: one flat rate
and the other one power-based. A flat fee, also referred to as a
flat rate or a linear rate, refers to a pricing structure that
charges a single fixed fee for a service, regardless of usage.
Then, the pricing function defined in equation (\ref{auth_payoff})
is $\mu(x)=1$. For Internet service providers, a flat rate pricing
scheme determines an access to the Internet for all customers of
the telco operator
 at a fixed tariff. Flat rate is common in broadband
access to the Internet in the USA and many other countries.
 In the power-based pricing scheme, we consider that the profit of the service provider comes from the network usage and it is proportional to the total power used by all the end users to send their traffic. Then, our model tends to limit the power of the end users as they have to pay the provider not based on the throughput they use but the power they consume. In this case, the service provider consider uses the following function $\mu(T_i)=T_i$. Note that introducing power based cost is common for CDMA
 \cite{ZSPB2011, AAG2010} and ALOHA networks (\cite{SE2009, GHAA2012}).

\section{General wireless model}\label{general}

We assume a standard wireless communication model as proposed in
\cite{Garg07}, where the signal received at the base station can
be written as $y=hx+z$, where $h$ stands for the block fading
process, $x$ the signal transmitted by the end user and $z$ is the
additive Gaussian noise. We assume coherent communication such
that the fading channel gain is constant over each block fading
length. Moreover, the additive Gaussian noise $z$ at the receiver
is i.i.d circularly symmetric and $z\sim\mathcal{CN}(0,\sigma^2)$.
Two wireless scenarios are considered: interference-free and
interference channel.

\subsection{Interference-free model}

We consider the situation where the end users have access to the
network interference from other end users, like in an Orthogonal
Frequency Division Multiplexing (OFDM) wireless technology in
which each end user has his own transmission frequency. So, there
is no interference between end users since the service provider
supplies different bandwidth frequency for each end user. Then,
the signal-to-noise ratio (SNR) of each end user $i$ is given by
$\gamma_i=\frac{LhT_i}{(W/n)\sigma^2}$ where $T_i$ is user's $i$
transmit power and $L$ is the crosstalk coefficient. We assume a
symmetric system such that all end users have the same channel
gain $h$. Moreover the service provider shares the leased
bandwidth frequency $W$ equally between the end users. Then we
assume that user $i$'s utility is given as follows:
\begin{eqnarray}
u_{i}(T_i)=(W/n)\ln\left(1+LhT_i/(W\sigma^2/n)\right)-C_P\mu(T_i).
\label{user1}
\end{eqnarray}
For the flat-rate pricing scheme, i.e. $\mu(x)=1$, it is natural
to assume that each end user is going to consume all the capacity.
Indeed the utility function $u_i$ is strictly increasing depending
on $T_i$. Then, each end user transmits with the maximal power
$\overline{T}$. Using this transmission power, each end user gets
a throughput equals to
$\frac{W}{n}\ln\left(1+\frac{Lh\overline{T}}{(W/n)\sigma^2}\right)$
and has to pay $C_P$ for the service. Then, each end user accepts
the provider's service if his utility (the difference between the
throughput and the charge) is non-negative. Indeed, there is no
incentive to transmit if his net utility is negative as it is
generally assume in economic models in telecommunication
\cite{Korcak12}. Moreover, since the service provider intends to
maximize his profit, then
\begin{equation}
\label{e_C_F}
C_P=(W/n)\ln\left(1+Lh\overline{T}/(W\sigma^2/n)\right).
\end{equation}
We have the first result supplying the solution of the multilevel
supply chain spectrum problem for flat-rate and power based
pricing schemes with interference-free model.

\begin{Theorem}\label{theo1} If there is no interference between end users, the equilibrium solution of the multilevel economic problem $(C_W,W,C_P,\bT)$ and corresponding payoffs of the service provider and the spectrum owner are given in Table~\ref{Tbl1}.

\begin{table}
\caption{\label{Tbl1} The optimal solution for interference-free
model}
%\begin{center}
\begin{tabular}{ccccc}
\hline\hline\noalign{\smallskip}
&Power based pricing&Flat rate  pricing\\
\noalign{\smallskip}\hline\hline\noalign{\smallskip}
$C_W$&$\frac{1}{4}$&$\approx 0.468$\\
$W$&$\frac{nLh\overline{T}}{\sigma^2}$&$\approx\frac{0.462nLh\overline{T}}{\sigma^2}$\\
$C_P$&$\frac{Lh}{2\sigma^2}$&$\approx\frac{0.532Lh\overline{T}}{\sigma^2}$\\
$\bT$&$(\overline{T},\ldots,\overline{T})$&$(\overline{T},\ldots,\overline{T})$\\
$v_P$&$\frac{nLh\overline{T}}{4\sigma^2}$&$\approx\frac{0.316nLh\overline{T}}{\sigma^2}$\\
$v_A$&$\frac{nLh\overline{T}}{4\sigma^2}$&$\approx\frac{0.216nLh\overline{T}}{\sigma^2}$\\
\noalign{\smallskip}\hline\hline
\end{tabular}
%\end{center}
\end{table}
\end{Theorem}

\noindent First, one important remark is that the optimal
frequency bandwidth tariffs $C_W^{PB-IF}=0.25$ and
$C_W^{FR-IF}=0.468$, for both pricing schemes,
 do not depend on the number $n$ of end users nor network's parameters. Here and below,
upper indexes {\it FR-IF} and {\it PB-IF} mean {\it flat rate} and {\it power
based} pricing schemes for the {\it interference-free} model
correspondingly. Moreover, power-based tariff is almost twice less
$C_W^{FR-IF}/C_W^{PB-IF}\approx 1.87$. Considering the power-based
pricing scheme the provider and the spectrum owner share profit
equally $v^{PB-IF}_A=v^{PB-IF}_P=\frac{\displaystyle nhL
\overline{T}}{\displaystyle 4\sigma^2}$ meanwhile, with the flat
rate pricing scheme, the service provider's payoff is essentially
greater than the spectrum owner's one, i.e.
$v^{PB-IF}_A/v^{PB-IF}_P=1.462$. Then, using a power-based pricing
scheme, the market is shared equally between the service provider
and the spectrum owner, whereas a flat-rate pricing scheme makes
the service provider's payoff higher than the spectrum owner's
payoff. Moreover, the total profit $v_P+v_A$ generated for both
decision makers, the service provider and the spectrum owner, is
higher with the flat rate pricing scheme, i.e.
$\frac{0.532nLh\overline{T}}{\sigma^2}$, compared to the total
profit with the power-based pricing scheme, i.e.
$\frac{0.5nLh\overline{T}}{\sigma^2}$. Finally, surprisingly, each
end user chooses also to transmit using maximum power, even if a
power-based pricing is used by the provider instead of a flat rate
pricing scheme. Thus, we can conclude that from one hand, in the
context of non-interfering end users, a power-based pricing scheme
will not help the spectrum owner to control the total power
consumption of the network but will permit for him to get more
market share and then earn more revenue from the spectrum rent.
From the other hand, the flat rate pricing scheme is preferable
for the service provider since it brings him higher profit
compared to using a power based pricing scheme which is moreover,
more difficult to build in practice.

\subsection{Interference channels}

In this section we consider that the access to the network is
performed considering user's interference. The signal to
interference and noise ratio (SINR) $\gamma_i$ of end user $i$ is
given by
$$
\gamma_i(\bT)=L h T_i \bigr/\left(W\sigma^2+h\sum_{j=1,j\not=i}^n
T_j\right).
$$
We consider a symmetric system, then the block fading process is the average one and it is the same for all users. So, the end user's $i$ payoff depends on the user action $T_i$ and
also on the actions of the other end users. We denote by $T_{-i}$
the vector of the actions of all the other players $j \neq i$. The
end user's $i$ payoff becomes
$$
u_{i}(\bT)=W\ln\left(1+\gamma_i(\bT)\right)-C_P\mu(T_i).
$$
For a fixed usage tariff $C_P$, the vector of end user's
strategies $\bT^*$ is a Nash
equilibrium (NE) if, by definition \cite{fudenberg91},
$$
\forall i \in \{1,\ldots,n\}, \quad \forall T_i\in
[0,\overline{T}], \quad u_{i}(T^*_i,\bT^*_{-i})\geq
u_{i}(T_i,\bT^*_{-i}),
$$
where $\bT_{-i}$ is the vector of all the end users instead
user $i$. \noindent We have the next result supplying the solution
of the multilevel supply chain spectrum problem for flat-rate and
power based pricing schemes with interfering end users. We take
into consideration then the NE between the end users for a given
decision of the service provider and the spectrum owner.

\begin{Theorem}\label{theo2} If there is interference
between end users, the solution of the multilevel economic problem
$(C_W,W,C_P,\bT)$ and corresponding payoffs $v_P$ and $v_A$
to the provider and the spectrum owner  are given in
Table~\ref{Tbl2}.

\begin{table*}[ht]
\begin{center}
\caption{\label{Tbl2} The optimal solution for interference
channels model}
\begin{tabular}{ccccc}
\hline
\hline\noalign{\smallskip}
&{\normalsize Power based pricing}&{\normalsize Flat rate  pricing}\\
\noalign{\smallskip}\hline\hline\noalign{\smallskip}
$W$&$\frac{(n+L-1)h\overline{T}}{\sigma^2}$&$W(C_W)=\arg\max_{W<\left(\frac{L}{\exp(C_W/n)-1}+1-n\right)\frac{h \bar{T}}{\sigma^2}}\left(n\ln
\left(1+\frac{Lh\overline{T}}{W\sigma^2+h(n-1)\overline{T}}\right)-C_W\right)W$\\
$C_W$&$\frac{nL}{4(n+L-1)}$&$C_W=\arg\max_{C_W<n\ln(1+L/(n-1))} C_W W(C_W)$\\
$C_P$&$\frac{Lh}{2\sigma^2}$&$W(C_W)\ln\left(1+\frac{Lh\overline{T}}{W(C_W)\sigma^2+(n-1)h\overline{T}}\right)$\\
$\textbf{T}$&$(\overline{T},\ldots,\overline{T})$&$(\overline{T},\ldots,\overline{T})$\\
$v_P$&$\frac{nLh\overline{T}}{4\sigma^2}$&$nW(C_W)\ln\left(1+\frac{Lh\overline{T}}{W(C_W)\sigma^2+(n-1)h\overline{T}}\right)-C_W\,W(C_W)$\\
$v_A$&$\frac{nLh\overline{T}}{4\sigma^2}$&$C_W\,W(C_W)$\\
\noalign{\smallskip}\hline\hline
\end{tabular}
\end{center}
\end{table*}
\end{Theorem}

For the power based pricing scheme in interference model the
optimal frequency bandwidth tariff $C_W$ is increasing on number
of users and the quality of network $L$. Specifically, this value
tends to $L/4$ when increasing the number of end users and
converges to $n/4$ when increasing the quality of network $L$.
Second, we observe also that the demand $W$ at the equilibrium
from the service provider in frequency bandwidth is increasing
depending on the number of end users $n$ to serve. Finally, the
network is over loaded such that the equilibrium user's strategies
$\textbf{T}$ is to use the maximum power. Then, it looks like
there is some kind of inside cooperative stimulus of all the
participants of the market (they just split the common profit) who
are in charge for network functionality and going to meet all the
user's demands. We conclude by observing the utility of each
decision maker at the equilibrium.

\begin{figure*}[ht]
\centering
\begin{tabular}{cccc}
\includegraphics[width=0.23\textwidth]{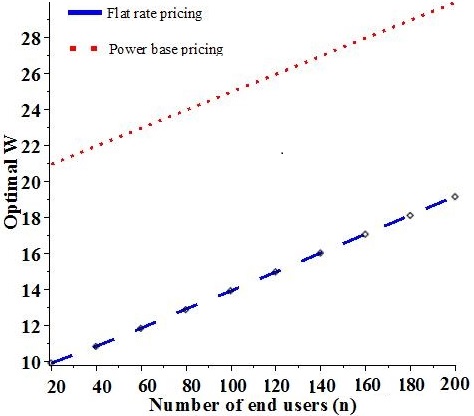} &
\includegraphics[width=0.23\textwidth]{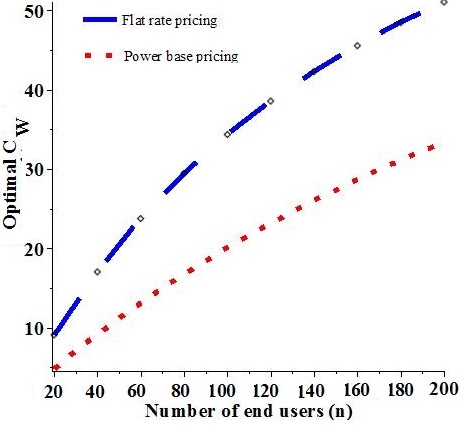} &
\includegraphics[width=0.23\textwidth]{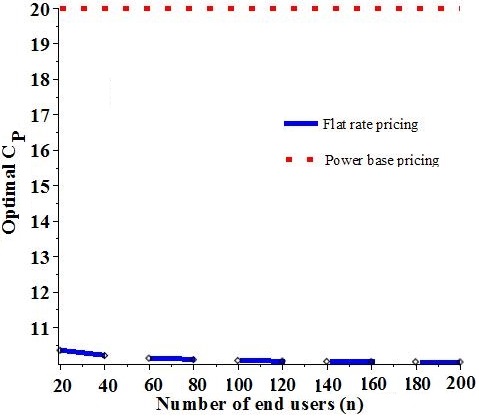} &
\includegraphics[width=0.23\textwidth]{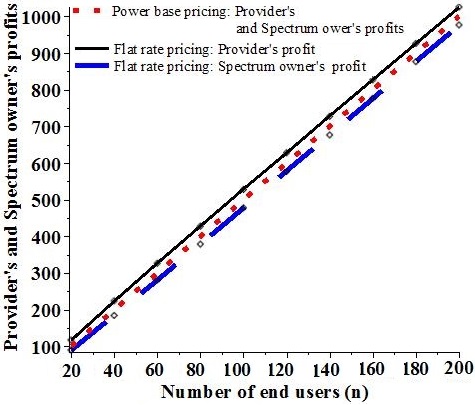}\\
(a)&(b)&(c)&(d)
\end{tabular}
\caption{\label{fig_1} (a) The optimal bandwidth $W$, (b) the optimal bandwidth cost $C_W$,  (c) the optimal transmission cost $C_P$ and  (d) the provider's and spectrum owner's
payoffs as functions on number of end users $n$ with $L= 400$, $h=1$, $\bar T = 0.5$ and $\sigma^2= 10.$}
\end{figure*}

\begin{figure*}[ht]
\centering
\begin{tabular}{cccc}
\includegraphics[width=0.22\textwidth]{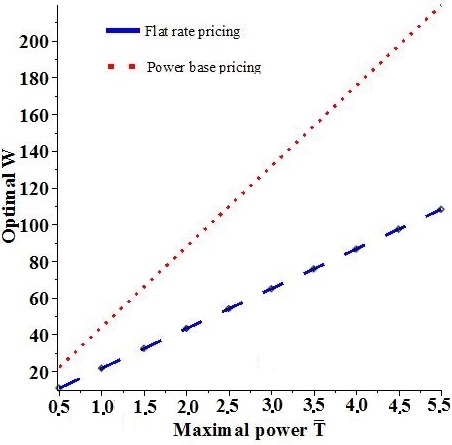} &
\includegraphics[width=0.22\textwidth]{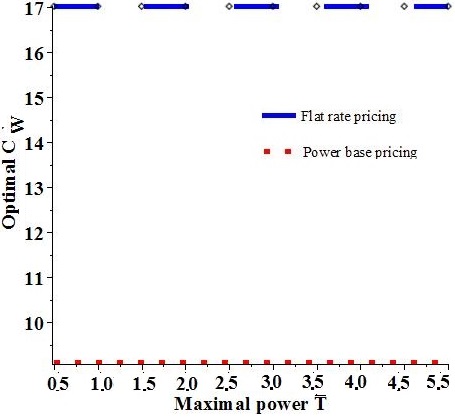} &
\includegraphics[width=0.22\textwidth]{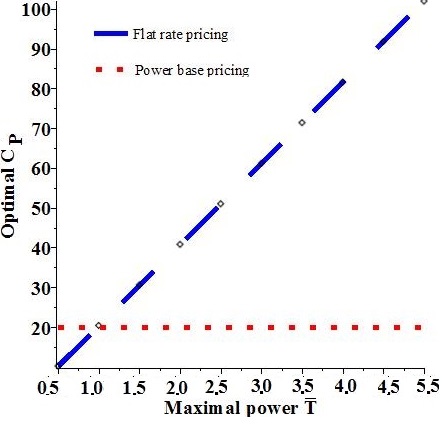} &
\includegraphics[width=0.21\textwidth]{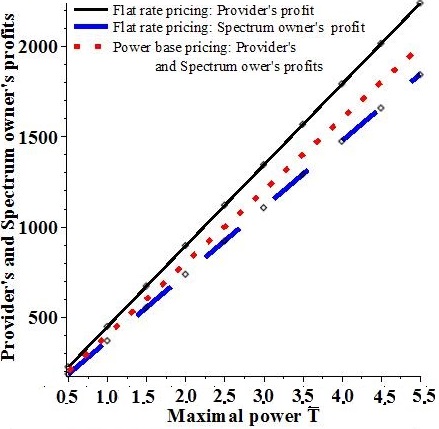}\\
(a)&(b)&(c)&(d)
\end{tabular}
\caption{\label{fig_2} (a) The optimal bandwidth $W$, (b) the optimal bandwidth cost $C_W$,  (c) the optimal transmission cost $C_P$ and   (d) the provider's and spectrum owner's
payoffs as functions on maximal power $\bar T$ with $L= 400$, $h=1$, $n = 40$ and $\sigma^2= 10.$}
\end{figure*}

\begin{itemize}
\item The profit of the spectrum owner and the profit of the service provider coincide at the equilibrium, namely,
$v_A=v_P=\frac{\displaystyle nhL \overline{T}}{\displaystyle
4\sigma^2}$. Thus, the spectrum owner assigns the bandwidth
frequency tariff in order to withdraw from the service provider a
fixed percent (namely, fifty percent) of his gross profit.
\item For each end user $i$, the optimal utility is given as follows
$$
u_{i}=\frac{\overline{T}h}{2\sigma^2}\left(
2(n+L-1)\ln\left(2\frac{n+L-1}{2n+L-2}\right)-L \right).
$$
which corresponds the following user's throughput
$$
R=\frac{\overline{T}h(n+L-1)}{\sigma^2}\ln\left(2\frac{n+L-1}{2n+L-2}\right).
$$
Thus,(since function $f(\xi)=\xi\ln(\xi/(\xi-L))$ is decreasing from infinity for $\xi\downarrow L$ to $L$ while $\xi$ tends to infinity)  the utility of each end user is strictly decreasing with $n$
and converges to 0 if $n$ tends to infinity. Moreover, the
throughput of each end user is decreasing with $n$ but converges
to the lower bound $\overline{T}hL/(2\sigma^2)$. Thus, whatever
the number of end users is, the throughput of each end used is
lower bounded, meaning that automatically a best-effort service
can be provided.

\item Figures~\ref{fig_1} and ~\ref{fig_2}  demonstrate that the most profitable pricing scheme for the
service provider is the flat rate pricing scheme when considering
end user's interference. It allows the spectrum owner to increase bandwidth cost (and it continuously  increasing up to its upper-bound $L/4$) While
 transmission cost  even drops down by the provider a bit for the flat rate pricing scheme, or
  it keeps on the same level ($Lh/(2\sigma^2)$) for power base pricing scheme. All together it leads to increasing in the provider's and spectrum owner's profits.
 Increasing consumption of the maximal power by end-users (which can be described as increasing intension of using the network by them) without increasing number of end users also allows the provider and spectrum owner to increase their profits, but they manage to get it  a bit different way. The spectrum owner'does not increase the bandwidth cost for both pricing schemes and gains just due to increasing demands in bandwidth. Meanwhile the provider to cover the extra expanses to buy more bandwidth has to lift up cost in flat rate scheme. In power based pricing scheme the extra income obtained due to more intensive using the network by end users allows the provider to buy more bandwidth for higher price without increasing transmission cost.
 \end{itemize}

\section{Special case: high SNR regime}\label{special}

\noindent A particular interesting wireless context is the high
SNR regime, i.e. for each end user $i$, $\gamma_i >>1$. This
assumption is always desirable in wireless networks (see
\cite{Ordonez07} for a recent paper on performance of multiplexing
MIMO systems in high-SNR regime) and leads to the following
relation $T_i >> \frac{W \sigma^2}{nLh}$ which gives a lower bound
on user's $i$ transmission power. Moreover, this assumption has
been also proposed in \cite{Duan11} for the study of an optimal
pricing mechanism for spectrum in a cognitive radio context. This
bound is relatively small as the number of end users $n$ is large
and then, this lower bound is realistic in a wireless network.
Thus, assuming high SNR regime, we have that $1+\gamma_i$ can be
approximated by $\gamma_i$. Then for the interference-free model,
the end user's $i$ payoff, assuming high SNR regime, is
approximated by the following expression
$$
u_i(\textbf{T})=\frac{W}{n}\ln\left(\frac{LhT_i}{(W/n)\sigma^2}\right)-C_P\mu_i(T_i)
$$
and for the interference user model it is given as follows
$$
u_{i}(\textbf{T})=W\ln\left(\frac{LhT_i}{W\sigma^2+h\sum_{j=1,j\not=i}^n
T_j}\right)-C_P\mu_i(T_i).
$$
To deal with high SNR approximation considering the power-based
pricing scheme, we assume that the bandwidth which the spectrum
owner can sell to the service provider is upper-bounded, namely,
that $W<\overline{W}$. Then we have the next result supplying the
solution of the multilevel supply chain spectrum game for
flat-rate and power based pricing schemes, and considering the
high SNR assumption.

\begin{Theorem}\label{theo3} Considering high SNR assumption, the equilibrium of the multilevel economic problem
$(C_W,W,C_P,\textbf{T})$ and corresponding payoffs $v_P$ and $v_A$
for the service provider and the spectrum owner are given in
Table~\ref{Tbl3} where $\epsilon$ is any enough small positive \footnote{The function $LambertW(.)$ is the {\it
LambertW} function satisfying the equation
$\mbox{LambertW}(x)e^{\mbox{LambertW}(x)}=x$.}.

\begin{center}
\begin{table*}[ht]
\begin{center}
\caption{\label{Tbl3} The equilibrium of the
multilevel spectrum game in high SNR regime.}
{\small
\begin{tabular}{ccccc}
\hline\hline
&Power based&Power based&Flat rate&Flat rate\\
&Interference-free&Interference&Interference-free&Interference\\\hline\hline
$C_W$&$1-\epsilon$&$n-\epsilon$&$1$&$n\mbox{LambertW}^2\left(\frac{n-1}{L}e^{\frac{n+C_W}{n}}\right)+C_W=n$\\
$W$&$\overline{W}$&$\overline{W}$&$\approx\frac{0.135nLh\overline{T}}{\sigma^2}$&$W=\frac{(n-1)h\overline{T}}{\sigma^2}\left(\frac{1}{\mbox{LambertW}\left(\frac{n-1}{L}e^{\frac{n+C_W}{n}}\right)}-1\right)$\\
$C_P$&$\overline{W}/(\overline{T}n)$&$\overline{W}/\overline{T}$&$\approx\frac{0.271Lh\overline{T}}{\sigma^2}$&$W\ln\left(\frac{Lh\overline{T}}{W\sigma^2+(n-1)h\overline{T}}\right)$\\
$v_P$&$\overline{W}\epsilon$&$\overline{W}\epsilon$&$\approx\frac{0.135nLh\overline{T}}{\sigma^2}$&$nW\ln\left(\frac{Lh\overline{T}}{W\sigma^2+(n-1)h\overline{T}}\right)-C_WW$\\
$v_A$&$\overline{W}(1-\epsilon)$&$\overline{W}(n-\epsilon)$&$\approx\frac{0.135nLh\overline{T}}{\sigma^2}$&$C_WW$\\
\hline\hline
\end{tabular}
}
\end{center}
\end{table*}
\end{center}
\end{Theorem}

It is quite interesting that considering the power-based pricing
scheme, the spectrum owner withdraws all the profit from the
service provider, just leaving him a bit, i.e.
$\overline{W}\epsilon$, to  keep working in the business. For the
flat rate pricing scheme, the spectrum owner and the service
provider share the total profit. Also, it is interesting to note
that for the interference-free model at high SNR regime, the
spectrum tariff $C_W^{FR-IF-H}$ of the spectrum owner is more than
twice higher than  the one in general regime $C_W^{FR-IF}$ since
$$
C_W^{FR-IF-H}/C_W^{FR-IF}=1/0.468=2.14.
$$
Meanwhile, when the end
users are going to exploit the network intensively, the tariff
applied by the service provider
$$
C_P^{FR-IF-H}/C_P^{FR-IF}=0.271/0.532=0.51
$$
is reduced two times.
The service provider's payoff reduces also since
$$
v_P^{FR-IF-H}/v_P^{FR-IF}=0.135/0.315=0.429.
$$
We can conclude
this remark by saying that if the service provider does not
control the number of end users and if this number becomes large,
using an OFDM technology the noise can be very low for each end
user and then his SNR very high. Thus, the provider's payoff is
divided by 2.

\section{Conclusions}\label{sec:conc}

In this paper, we have studied a multilevel economic model of
wireless networks with two pricing schemes: a flat-rate or a
power-based. The different decision makers are the end users, a
service provider and a spectrum owner. The proposed economic model
can be applied in several networking contexts like cognitive radio
network, where a company or an authority which has the control on
the resource (the spectrum here) rents frequencies to service
providers and at the end of this spectrum supply chain, end users
pay for their services to the service providers. This multilevel
model gives interesting results on how optimal economic relations
are built between the different decision makers. In particular,
for the power-based pricing scheme, the service provider and the
spectrum owner split equally the total market profit. Whereas,
considering a flat-rate pricing scheme it is more profitable for
the service provider. Of course this conclusion should be taken
with caution, and that a more complete model is needed to make
stronger conclusions. In terms of quality of service, the
power-based tariff at the equilibrium depends on the quality of
the network and not on the number of end users. Another important
result deals with the flat-rate pricing scheme. Considering this
scheme, if the end users are going to exploit the network
intensively then, it leads to double increasing of the bandwidth
frequency tariff by the spectrum owner. It leads also to a bit
less than double increasing of the flat rate tariff to the end
users applied by the service provider. Also,  we show that
power-based pricing leads to an equal profit split between
spectrum owner and provider, and that the power-based pricing can
lead to the spectrum owner getting more profit than a flat-pricing
scheme even though flat-pricing is better for extracting the
surplus from the users. In future works, we plan to investigate
that kind of multilevel model for  a important new wireless market
by considering Cognitive Radio Networks in which second hand
wireless providers can sell spectrum holes to secondary users.

\bibliographystyle{ieeetr}
\bibliography{ValueTools2013_for_archive}

  % sigproc.bib is the name of the Bibliography in this case
% ACM needs 'a single self-contained file'!
%\balancecolumns
\section{Appendix}
%\section{Appendix}
%\label{sec:Appendix}
%%%%%%%%%%%%%%%%%%%%%%%%%%%%%%%%%%%%%%%%%%%%%%%%%%%%%%%%%%%
\subsection{Proof of Theorem 1}
%%%%%%%%%%%%%%%%%%%%%%%%%%%%%%%%%%%%%%%%%%%%%%%%%%%%%%%%%%%%

\indent
{\it We consider the flat rate pricing scheme, i.e.
$\mu(x)\equiv 1$.} Then, the utility of each end users is strictly
increasing with the transmission power $T$, and thus each end will
use the maximum power $\overline{T}$. Moreover, each end-user
accepts the provider's service if and only if his utility is
positive, which gives the maximum tariff an end-user is able to
pay:
\small
$$
C_P=\frac{W}{n}\ln\left(1+\frac{Lh\overline{T}}{(W/n)\sigma^2}\right).
$$
\normalsize \noindent
Given this, the provider's payoff turns into the following
function:
\small
$$
v_P(C_P,W,C_W)=W\ln\left(1+\frac{nh\overline{T}L}{W\sigma^2}\right)-C_WW.
$$
\normalsize \noindent
Note that this function does not depend on $C_P$ and:
\small
$$
\frac{\partial v_P}{\partial W}(C_P,W,C_W)=\ln\left(1+\frac{L
h\overline{T}n}{W\sigma^2}\right)-\frac{Lh\overline{T}n}{Lh\overline{T}n+W\sigma^2}-C_W.
$$
\normalsize \noindent
We look now for the optimal quantity of bandwidth the provider
will rent from  the spectrum provider  depending on the tariff
$C_W$. The optimal frequency bandwidth to lease is given as
follows:
\small
\begin{equation}
\label{eW}
W=W(C_W):=-\frac{(Lh\overline{T}n/\sigma^2)\mbox{LambertW}(-e^{-1-C_W})}{1+\mbox{LambertW}(-e^{-1-C_W})}.
\end{equation}
\normalsize \noindent Finally, the spectrum owner's payoff can be rewritten
depending only on $C_W$ as follows:
\small
\begin{equation*}
\begin{split}
v_A(C_W)=-\frac{(Lh\overline{T}n/\sigma^2)\mbox{LambertW}(-e^{-1-C_W})}{1+\mbox{LambertW}(-e^{-1-C_W})}C_W.
\end{split}
\end{equation*}
\normalsize \noindent
Note that the derivative of this function is:
\small
\begin{equation*}
\begin{split}
\frac{d v_A(C_W)}{d
C_W}&=(Lh\overline{T}n/\sigma^2)\mbox{LambertW}\left(-e^{-1-C_W}\right)\\
&\times\frac{C_W-\left(1+\mbox{LambertW}\left(-e^{-1-C_W}\right)\right)^2}{(1+\mbox{LambertW}(-e^{-1-C_W}))^3}.
\end{split}
\end{equation*}
\normalsize \noindent
So, the optimal bandwidth frequency tariff is given as the root of the equation
$$
C_W-\left(1+\mbox{LambertW}\left(-e^{-1-C_W}\right)\right)^2=0.
$$
and then since $C_W\approx 0.468$ the result  follows for the flat
rate pricing scheme.

%%%%%%%%%%%%%%%%%%%%%%%%%%%%%%%%%%%%%%%%%%%%%%%%%%%%%%%%%%%%%%%%%%%%%%%%%%%%%%%%

{\it We consider the power based  pricing scheme, i.e.
$\mu(x) \equiv x$.}  {\bf  At the first step} of the multilevel game, we have to
find the optimal end user strategies. To do so, we first determine
the optimal transmission power of each end user $i$ by computing
the derivative of his utility function:
\small
$$
\frac{du_{i}}{d T_i}(T_i)=\frac{\displaystyle WL h }{\displaystyle
W\sigma^2+LhnT_i}-C_P.
$$
\normalsize \noindent
Thus, the optimal transmission power of user $i$ is
$T_i=T(C_P,W)$:

\small
\begin{equation}
\label{e_W_000_a} T(C_P,W)=
  \begin{cases}
    \overline{T}, & \displaystyle C_P\leq \frac{\displaystyle LWh}{\displaystyle W\sigma^2+Lhn\overline{T}},\\
\frac{\displaystyle W}{\displaystyle n}\left(\frac{\displaystyle
1}{\displaystyle C_P}-\frac{\displaystyle \sigma^2}{\displaystyle
Lh}\right) , &
\frac{\displaystyle LWh}{\displaystyle W\sigma^2+Lhn\overline{T}}<C_P<\frac{\displaystyle Lh}{\displaystyle\sigma^2},\\
0,& \frac{\displaystyle Lh}{\displaystyle\sigma^2}\leq C_P.
  \end{cases}
\end{equation}
\normalsize \noindent  {\bf  Now we consider the second step} where
the provider, knowing the equilibrium user's strategies
(\ref{e_W_000_a}) have to decide which tariff $C_P$ to assign and
how much bandwidth $W$ to lease to maximize his revenue. The
provider's payoff at the second step of the game is given by:
\small
\begin{equation}
\label{e_W0_a}
\begin{split}
v_P&(C_P,W)
  =-C_W W\\
  &+\begin{cases}
    n\overline{T}C_P, & \displaystyle C_P\leq \frac{\displaystyle LWh}{\displaystyle W\sigma^2+Lhn\overline{T}},\\
W\left(1-\frac{\displaystyle \sigma^2 C_P}{\displaystyle
Lh}\right), &
\frac{\displaystyle LWh}{\displaystyle W\sigma^2+Lhn\overline{T}}<C_P<\frac{\displaystyle Lh}{\displaystyle\sigma^2},\\
0,& \frac{\displaystyle Lh}{\displaystyle\sigma^2}\leq C_P.
  \end{cases}
\end{split}
\end{equation}

\normalsize \noindent Then, the provider's revenue achieves its
maximum value when the tariff for the users $C_P$ is such that
\small
\begin{equation}
\label{e_W1_a} C_P=C_P(W):=\frac{\displaystyle LWh}{\displaystyle
W\sigma^2+Lhn\overline{T}}.
\end{equation}
\normalsize \noindent
Substituting (\ref{e_W1_a}) into (\ref{e_W0_a}) leads to the
following expression of the provider's revenue depending only on
the quantity of bandwidth $W$ to lease:
\small
\begin{equation}
\label{p1_a}
\begin{split}
v_P(C_P,W)=v_P(W):=\frac{\displaystyle
n\overline{T}LWh}{\displaystyle W\sigma^2+Lhn\overline{T}}-C_WW.
\end{split}
\end{equation}
\noindent The derivative of this expression depending on $W$ is:
\small
\begin{equation*}
\begin{split}
\frac{d
v_P}{dW}(W)&=\frac{(Lnh\overline{T})^2(1-C_W)-2Lnh\overline{T}C_WW-C_W\sigma^4W^2}{\left(W\sigma^2+Lnh\overline{T}\right)^2}.
\end{split}
\end{equation*}
\normalsize Then,

\noindent
(a) if
$C_W\geq 1$ then $v_P(W)$ is decreasing for positive $W$,

\noindent
(b) if
$C_W <1$ then $v_P(W)$ is increasing on the interval
$\left(0,
(1-\sqrt{C_W})\overline{T}hnL/(
\sqrt{C_W}\sigma^2)\right)$ and it is decreasing on the interval
$\left(
(1-\sqrt{C_W})\overline{T}hnL/(
\sqrt{C_W}\sigma^2),\infty\right)$.\normalsize

\noindent Thus, the optimal bandwidth $W=W(C_W)$ which maximizes
the provider's revenue is given as follows:

\small
\begin{equation}
\label{e_W20} W(C_W)=
\begin{cases}
0, & C_W\geq 1, \\
\frac{\displaystyle (1-\sqrt{C_W})\overline{T}hnL}{\displaystyle
\sqrt{C_W}\sigma^2},& C_W<1.
\end{cases}
\end{equation}

\normalsize \noindent Then,  {\bf  at the third and last step of the
game}, taking into account (\ref{e_W20})  the spectrum owner's
payoff is given by
\small
\begin{equation*}
\begin{split}
v_A(C_W)=\begin{cases}
0, & C_W\geq 1, \\
\frac{\displaystyle
\sqrt{C_W}(1-\sqrt{C_W})\overline{T}hnL}{\displaystyle \sigma^2},& C_W<1.
\end{cases}
\end{split}
\end{equation*}
\normalsize \noindent Then, the optimal spectrum owner's strategy
maximizing $v_A(C_W)$ is to assign the frequency bandwidth tariff
$C_W=1/4$, and the result follows. \cqfd

%%%%%%%%%%%%%%%%%%%%%%%%%%%%%%%%%%%%%%%%%%%%%%%%%%%%%%%%%%%%%%%%%%%%%%%%%%%%%%%%%%
%%% Theorem 2
%%%%%%%%%%%%%%%%%%%%%%%%%%%%%%%%%%%%%%%%%%%%%%%%%%%%%%%%%%%%%%%%%%%%%%%%%%%%%%%%

%%%%%%%%%%%%%%%%%%%%%%%%%%%%%%%%%%%%%%%%%%%%%%%%%%%%%%%%%%%
\subsection{Proof of Theorem 2}
%%%%%%%%%%%%%%%%%%%%%%%%%%%%%%%%%%%%%%%%%%%%%%%%%%%%%%%%%%%%

{\it We consider the flat rate pricing scheme, i.e.
$\mu(x)\equiv 1$.} It is clear that the end users optimal behaviour is
$\textbf{T}=(\overline{T},\ldots,\overline{T})$. Then, for the
provider, the tariff in order to maximize his payoff is:
$$
C_P=C_P(W):=W\ln
\left(1+\frac{Lh\overline{T}}{W\sigma^2+h(n-1)\overline{T}}\right).
$$
Thus, the provider's payoff at the NE of the end users becomes:
\small
\begin{equation*}
\begin{split}
v_P(C_P,W,C_W)&=v_P(C_P(W),W,C_W)=nC_P-WC_W\\
&=\left(n\ln
\left(1+\frac{Lh\overline{T}}{W\sigma^2+h(n-1)\overline{T}}\right)-C_W\right)W.
\end{split}
\end{equation*}
\normalsize
Then,
\small
$$\
\arg\limits_W\max v_P(C_P(W),W,C_W)
\begin{cases}
=0,& C_W\geq n\ln\left(1+\frac{\displaystyle L}{\displaystyle  n-1}\right),\\
>0,& C_W<n\ln\left(1+\frac{\displaystyle  L}{\displaystyle  n-1}\right),
\end{cases}
$$
\normalsize
\noindent
and the result follows.

{\it We consider the power based pricing scheme, i.e.
$\mu(x)\equiv x$.} As usual in hierarchical optimization problems, in
order to compute a solution, we consider the optimization problem
starting from the bottom optimization level (the end users
equilibrium) to the top level optimization problem (the spectrum
owner).

{\bf At the first step}  we have to find the Nash equilibrium of
the non-cooperative power control game between the end users. The
utility function of end user $i$ is given by:
\small
$$
u_{i}(\textbf{T})=W\ln\left(1+\frac{LhT_i}{W\sigma^2+h\sum_{j=1,j\not=i}^n
T_j}\right)-C_PT_i.
$$
\normalsize \noindent
In order to obtain the Nash equilibrium, we determine the
best-response strategy of each end user by computing the following
partial derivative:
\small
$$
\frac{\partial u_{i}}{\partial
T_i}(\textbf{T})=\frac{\displaystyle WL h }{\displaystyle
W\sigma^2+LhT_i+h\sum_{j=1,j\not=i}^n T_j}-C_P.
$$
\normalsize \noindent
Thus, as the game is symmetric, the equilibrium strategy
$T_i=T(C_P,W)$ for each user $i$ is
\small
\begin{equation}
\label{e_W_000} T(C_P,W)=
  \begin{cases}
    \overline{T}, & \displaystyle C_P\leq \frac{\displaystyle LWh}{\displaystyle W\sigma^2+(n-1+L)h\overline{T}},\\
\frac{\displaystyle\left(\frac{\displaystyle Lh}{\displaystyle
C_P}-\sigma^2\right) W}{\displaystyle h(n-1+L)}, &
\frac{\displaystyle LWh}{\displaystyle
W\sigma^2+(n-1+L)h\overline{T}}<C_P\\
&\mbox{and } C_P<\frac{\displaystyle Lh}{\displaystyle\sigma^2},\\
0,& \frac{\displaystyle Lh}{\displaystyle\sigma^2}\leq C_P.
  \end{cases}
\end{equation}
\normalsize

{\bf At the second step}, the provider,
knowing the equilibrium user's strategies (\ref{e_W_000}), has to
determine which tariff $C_P$ to assign and how much bandwidth $W$
to lease in order to maximize his revenue. The provider's payoff
at the second step of the game is given by:
\small
\begin{equation}
\label{e_W0}
\begin{split}
&v_P(C_P,W)=-C_W W\\
&+\begin{cases}
    n\overline{T}C_P, & \displaystyle C_P\leq \frac{\displaystyle LWh}{\displaystyle W\sigma^2+(n-1+L)h\overline{T}},\\
\frac{\displaystyle\left(Lh-\sigma^2C_P\right) nW}{\displaystyle
h(n-1+L)}, & \frac{\displaystyle LWh}{\displaystyle
W\sigma^2+(n-1+L)h\overline{T}}<C_P\\
&\mbox{and } C_P<\frac{\displaystyle Lh}{\displaystyle\sigma^2},\\
0,& \frac{\displaystyle Lh}{\displaystyle\sigma^2}\leq C_P.
  \end{cases}
\end{split}
\end{equation}
\normalsize
\noindent Then, the provider revenue $v_P(C_P,W)$ is
continuous and piece-monotonous on $C_P$ such way that

\noindent
(a) $v_P(C_P,W)$ is increasing for $C_P\leq \frac{ LWh}{W\sigma^2+(n-1+L)h\overline{T}},$

\noindent
(b) $v_P(C_P,W)$ is decreasing for $\frac{ LWh}{
W\sigma^2+(n-1+L)h\overline{T}}<C_P<\frac{ Lh}{\sigma^2}$,

\noindent(c) $v_P(C_P,W)$ is constant for $\frac{
Lh}{\sigma^2}\leq C_P$.

Then, the provider revenue achieves its optimum value when the
tariff for the users $C_P$ is such that
\small
\begin{equation}
\label{e_W1} C_P=C_P(W):= LWh/\left(
W\sigma^2+(n-1+L)h\overline{T}\right).
\end{equation}
\normalsize \noindent
Substituting (\ref{e_W1}) into (\ref{e_W0}) leads to the following
expression of the provider's revenue depending only on the
quantity of bandwidth $W$ to lease:

\small
\begin{equation}
\label{p1}
\begin{split}
v_P(W):=v_P(C_P(W),W)=\frac{\displaystyle
nLWh\overline{T}}{\displaystyle
W\sigma^2+(n-1+L)h\overline{T}}-C_WW.
\end{split}
\end{equation}
\normalsize

\noindent Note that the derivative depending on $W$ of the
provider's revenue is:
\small
\begin{equation*}
\begin{split}
\frac{d
v_P}{dW}(W)=\frac{A_2W^2+A_1W+A_0}{(W\sigma^2+h(n+L-1)\overline{T})^2},
\end{split}
\end{equation*}
\normalsize \noindent
with
\small
\begin{equation*}
\begin{split}
A_0&:=h^2\overline{T}^2(n+L-1)(nL-C_W(n+L-1)),\\
A_1&:=-2C_W\sigma^2h\overline{T}(n+L-1),\\
A_2&:=-C_W\sigma^2.
\end{split}
\end{equation*}
\normalsize \noindent
The following
equation $A_2W^2+A_1W+A_0=0$ has the two roots:
\small
$$
W_\pm=(-C_W(n+L-1)\pm
\sqrt{C_WnL(n+L-1)})\frac{\overline{T}h}{C_W\sigma^2}.
$$
\normalsize
\noindent
Then we have the following cases:
\begin{description}
\item[(a)] if
$C_W\geq \frac{nL}{L+n-1}$ then $v_P(W)$ is decreasing for
positive $W$,
\item[(b)] if
$C_W <\frac{nL}{L+n-1}$ then $v_P(W)$ is increasing for
$W<W_+$ and
decreasing for $W>W_+$.
\end{description}
\noindent Thus, the optimal bandwidth $W=W(C_W)$ which maximizes
the provider's revenue is given as follows:
\small
\begin{equation}
\label{e_W2} W(C_W)=
\begin{cases}
0, & C_W\geq \frac{\displaystyle nL}{\displaystyle L+n-1}, \\
 (\sqrt{nL}-\sqrt{C_WnL(n+L-1)})\\
\times\frac{\displaystyle \sqrt{(n+L-1)}\overline{T}h}{\displaystyle
\sqrt{C_W}\sigma^2},&
\mbox{otherwise}.
\end{cases}
\end{equation}
\normalsize
\noindent  {\bf  Finally, at the third and last step of the
game}, taking into account (\ref{e_W2}), the spectrum owner's
payoff is given by:
\small
\begin{equation*}
\begin{split}
v_A(C_W)=\begin{cases}
0, & C_W\geq \frac{\displaystyle nL}{\displaystyle L+n-1}, \\
\frac{\displaystyle \sqrt{C_W(n+L-1)}\overline{T}h}{\displaystyle
\sigma^2}\\
\times(\sqrt{nL}-\sqrt{C_WnL(n+L-1)}),&
\mbox{otherwise}.
\end{cases}
\end{split}
\end{equation*}
\normalsize \noindent
If $C_W<\frac{\displaystyle nL}{\displaystyle L+n-1}$ we have that
\small
$$
\frac{d
v_A}{dC_W}(C_W)=\frac{\overline{T}h\sqrt{n+L-1}}{2\sigma^2\sqrt{C_W}}(\sqrt{nL}-2\sqrt{C_W(n+L-1)}).
$$
\normalsize \noindent Then, the optimal spectrum owner's strategy
is to assign the frequency bandwidth tariff $C_W$ as follows:
$C_W=\frac{nL}{4(n+L-1)}$ and the result follows. \cqfd

%%%%%%%%%%%%%%%%%%%%%%%%%%%%%%%%%%%%%%%%%%%%%%%%%%%%%%%%%%%%%%%%%%%%%%%%%%%%%%%%
%%%%%%%% Theorem 3
%%%%%%%%%%%%%%%%%%%%%%%%%%%%%%%%%%%%%%%%%%%%%%%%%%%%%%%%%

\subsection{Proof of Theorem 3}
%%%%%%%%%%%%%%%%%%%%%%%%%%%%%%%%%%%%%%%%%%%%%%%%%%%%%%%%%%%%
%%%%%%%%%%%%%%%%%%%%%%%%%%%%%%%%%%%%%%%%%%%%%%%%%%%%%%%%%%%%
{\it We consider the flat rate pricing scheme, i.e.
$\mu(x)=1$ in user-free model.} Considering the high SNR regime, we can make the following
approximation $\log(1+\gamma_i)=\log(\gamma_i)$. Then the
provider's payoff becomes:
$$
v_P(C_P,W,C_W)=W\ln
\left(\frac{Lnh\overline{T}}{W\sigma^2}\right)-WC_W.
$$
Note that the derivative of the provider's payoff is:
\begin{equation*}
\begin{split}
\frac{\partial v_P}{\partial W}(C_P,W,C_W)=\ln\left(\frac{L
hn\overline{T}}{W\sigma^2}\right)-1-C_W.
\end{split}
\end{equation*}
Thus, the optimal frequency bandwidth $W=W(C_W)$ to lease is
\begin{equation*}
W(C_W):=\frac{L hn\overline{T}}{\sigma^2e^{1+C_W}}.
\end{equation*}
\normalsize \noindent The spectrum owner's payoff turns into the
following expression:
\begin{equation*}
\begin{split}
v_A(C_W)&=C_W W(C_W)=\frac{L
hn\overline{T}C_W}{\sigma^2e^{1+C_W}}.
\end{split}
\end{equation*}
\normalsize \noindent Note that the derivative of this function
is:
\begin{equation*}
\begin{split}
\frac{d v_A}{d C_W}(C_W)=\frac{L
hn\overline{T}(1-C_W)}{\sigma^2e^{1+C_W}}.
\end{split}
\end{equation*}
\normalsize \noindent Then, the optimal bandwidth frequency tariff
is $C_W=1$ and the optimal quantity of bandwidth rent by the
provider is:
$$
W=\frac{L hn\overline{T}}{\sigma^2e^{2}}
$$
and the result follows:

{\it We consider the flat rate pricing scheme, i.e.
$\mu(x)=1$ in interference model.} Given the approximation of the SINR of each end users, the
provider's revenue is given by:
$$
v_P(C_P,W,C_W)=nW\ln
\left(\frac{Lh\overline{T}}{W\sigma^2+h(n-1)\overline{T}}\right)-WC_W.
$$
Note that the derivative of this expression depending on the
quantity of bandwidth $W$ is:

\begin{equation*}
\begin{split}
\frac{\partial v_P}{\partial W}(C_P,W,C_W)&=n\ln\left(\frac{L
h\overline{T}}{W\sigma^2+(n-1)h\overline{T}}\right)\\
&-\frac{W\sigma^2n}{W\sigma^2+(n-1)h\overline{T}}-C_W.
\end{split}
\end{equation*}
Thus, the optimal frequency bandwidth to lease is $W=W(C_W)$ with
\begin{equation*}
W(C_W):=\frac{(n-1)h\overline{T}}{\sigma^2}\left(\frac{\displaystyle
1}{\displaystyle\mbox{LambertW}\left(\frac{\displaystyle
n-1}{\displaystyle L}e^{\frac{\displaystyle n+C_W}{\displaystyle
 n}}\right)}-1\right).
\end{equation*}
\normalsize \noindent Given the optimal reaction of the provider,
the spectrum owner's payoff can be rewritten by:
\small
\begin{equation*}
\begin{split}
v_A(C_W)=\frac{(n-1)h\overline{T}}{\sigma^2}\left(\frac{\displaystyle
1}{\displaystyle\mbox{LambertW}\left(\frac{\displaystyle
n-1}{\displaystyle L}e^{\frac{\displaystyle n+C_W}{\displaystyle
 n}}\right)}-1\right)C_W.
\end{split}
\end{equation*}
\normalsize \noindent We compute the derivative of this function:
\small
\begin{equation*}
\begin{split}
\frac{d v_A(C_W)}{d
C_W}&=\frac{1-\mbox{LambertW}^2
\left(\frac{\displaystyle
n-1}{\displaystyle L}e^{\frac{\displaystyle n+C_W}{\displaystyle
n}}\right)-C_W}{\mbox{LambertW}\left(\frac{\displaystyle
n-1}{\displaystyle L}e^{\frac{\displaystyle n+C_W}{\displaystyle
n}}\right)}\\
&\times\frac{
\displaystyle (n-1)h\overline{T}/\sigma^2}{1+\mbox{LambertW}\left(\frac{\displaystyle
n-1}{\displaystyle L}e^{\frac{\displaystyle n+C_W}{\displaystyle
n}}\right)}.
\end{split}
\end{equation*}
\normalsize \noindent So, the optimal bandwidth frequency tariff
is given as the root of the equation
$$
\mbox{LambertW}^2\left(\frac{\displaystyle n-1}{\displaystyle
L}e^{\frac{\displaystyle n+C_W}{\displaystyle n}}\right)=1-C_W.
$$
and the result follows.

{\it We consider the power based pricing scheme, i.e.
$\mu(x)=x$ in interference model.} First we have to find the Nash
equilibrium of the non-cooperative power control game between the
end users. The utility function of end user $i$ is given by:
$$
u_{i}(\bT)=W\ln\left(\frac{LhT_i}{W\sigma^2+h\sum_{j=1,j\not=i}^n
T_j}\right)-C_PT_i.
$$
In order to obtain the Nash equilibrium, we determine the
best-response strategy of each end user by computing the following
partial derivative:
$$
\frac{\partial u_{i}}{\partial
T_i}(\textbf{T})=\frac{\displaystyle W }{\displaystyle T_i}-C_P.
$$
Thus, the user's Nash equilibrium is
$(\overline{T},\ldots,\overline{T})$ and the optimal tariff is
$C_P=C_P(W)=W/\overline{T}$. Thus, the payoffs to the provider and
spectrum owner turn into
$v_P(W,C_W)=(n-C_W)W$, $v_A(W,C_W)=C_WW$,
and the result follows.

{\it We consider the power based pricing scheme, i.e.
$\mu(x)=x$ in user-free model.} First we have to find the Nash
equilibrium of the non-cooperative power control game between the
end users. The utility function of end user $i$ is given by:
$$
u_{i}(\textbf{T})=\frac{W}{n}\ln\left(\frac{n
LhT_i}{W\sigma^2}\right)-C_PT_i.
$$
In order to obtain the Nash equilibrium, we determine the
best-response strategy of each end user by computing the following
partial derivative:
$$
\frac{\partial u_{i}}{\partial
T_i}(\textbf{T})=\frac{\displaystyle W }{\displaystyle n T_i}-C_P.
$$
Thus, the user's Nash equilibrium is
$(\overline{T},\ldots,\overline{T})$ and the optimal tariff is
$C_P=C_P(W)=W/\overline{T}$. Thus, the payoffs to the provider and
spectrum owner turn into $v_P(W,C_W)=(1-C_W)W$, $v_A(W,C_W)=C_WW$,
and the result follows. \cqfd

\end{document}